\documentclass[a4paper,12pt]{article}
\usepackage[polish, english]{babel}
\usepackage[cp1250]{inputenc}
\usepackage{amsmath,amsmath,graphicx,epsf,pstricks,amsfonts,epsfig,multicol}
\title{\textbf{Induced gravity and gauge interactions revisited}}
\author{Bogus{\l}aw
Broda\footnote{bobroda@uni.lodz.pl}\; and Micha{\l} Szanecki\footnote{michalszanecki@wp.pl}\\
\small\textit{Department of Theoretical Physics,}
\small\textit{University of {\L}\'od\'z}\\
\small\textit{Pomorska 149/153,\;90-236 {\L}\'od\'z, Poland}\\
}
\begin{document}
\maketitle
\begin{abstract}
\noindent
\begin{quote}
It has been shown that the primary, old-fashioned idea of
Sakharov's induced gravity and gauge interactions, in the
``one-loop dominance'' version, works astonishingly well yielding
phenomenologically reasonable results. As a byproduct, the issue
of the role of the UV cutoff in the context of the induced gravity
has been reexamined (an idea of self-cutoff induced gravity). As
an additional check, the black hole entropy has been used in the
place of the action. Finally, it has been explicitly shown that
the induced coupling constants of gauge interactions of the
standard model assume qualitatively realistic values.\\\\
\textbf{PACS number(s):} 11.15.Tk Other nonperturbative
techniques; 04.62.+v Quantum fields in curved spacetime; 04.70.Dy
Quantum aspects of black holes, evaporation, thermodynamics;
12.10.Dm Unified theories and models of strong and electroweak
interactions.
\end{quote}
\end{abstract}
\eject
\begin{center}\section{Introduction}\end{center}
\noindent The idea that fundamental interactions might be not so
fundamental as they appear, but induced by quantum fluctuations of
the vacuum emerges from the fifties of the 20th century. In 1967
Sakharov published his famous short paper on induced gravity
\cite{Sakharov}, and in the same year Zel'dovich presented a
parallel result concerning electrodynamics \cite{Zeldovich}. The
both authors have made use of some earlier observations coming
from papers (cited by them) by Landau and his collaborators. In
fact, the induced gauge theory was initiated four years earlier in
\cite{Bjorken}, and next followed by many people (see, e.g.\
\cite{Birula}). It was applied to the standard model in
\cite{Terazawa1}, whereas field theoretical realizations and
calculations concerning induced gravity were given in
\cite{Akama1},\cite{Akama2}. A logarithmic relation between the
gauge coupling constants and the Newton gravitational constant has
been noticed in \cite{Terazawa2} (this relation has been derived
in \cite{Landau} using another requirement). Some further, related
aspects have been elaborated in \cite{Akama3},\cite{Akama4}. It
seems that successful application of the idea of quantum vacuum
induced interactions to the two fundamental interactions
subsequently renewed interest in this subject. Actually, at
present, the very idea lacks a clear theoretical interpretation.
It can be treated either as an interesting curiosity or as an
unexplained deeper phenomenon. Anyway, coincidences are striking.
Our point of view is purely pragmatical, i.e.\ we claim that the
idea of quantum induced interactions does work.

The aim of our paper is to show that the idea of induced gauge
interactions (including gravity and, possibly, dark energy) in its
primary, old-fashioned, Sakharov's version yields
phenomenologically very realistic results. The phrase ``primary,
old-fashioned, Sakharov's version'' means ``one-loop dominance''
interpretation in the terminology of a review paper on Sakharov's
induced gravity \cite{Visser}. This standpoint assumes that at the
beginning there are no classical terms for gauge fields and
gravity. There are only (fundamental) matter fields present in the
classical action, and they are coupled to external gauge fields
and gravity. (The superior role of the matter fields awaits an
explanation in this framework.) Interestingly, and it was primary
inspiration, it appears that low-order one-loop calculations yield
proper classical terms for gauge and gravitational fields. Just
only this fact, akin to renormalizability, is by no means
surprising. But what is really surprising is that not only
appropriate functional terms emerge from these one-loop matter
field calculations but phenomenologically realistic numeric
coefficients as well.

As far as a conceptual side of the idea is concerned, in the case
of gravity, we have also proposed an alternative point of view.
Actually, there is some logical gap in the standard approach which
consists in imposing the Planck cutoff to derive the strength of
gravitational interactions which subsequently yields the Planck
cutoff itself. Namely, we propose to shift the focus to analyzing
the role of relation between the ``Schwarzschild'' radius and the
mass, leaving the Newton gravitational constant undetermined
(self-cutoff induced gravity). We have also suggested to use the
entropy instead of the action as an independent check of the whole
procedure. Finally, we have explicitly estimated coupling
constants of fundamental interactions, which appear to assume
realistic values. In our paper, gravity (possibly, including dark
energy) and gauge interactions are treated uniformly, i.e.\ the
both kinds of interactions are analysed parallelly and the both
kinds of interactions are approached in the framework of the same
and very convenient method: Schwinger's proper time and the
Seeley--DeWitt heat-kernel expansion.
One should mention that the natural idea of using the heat-kernel
expansion (in the euclidean version) to get the gravitational
induced action was introduced in \cite{DS1}, whereas temperature
dependence of the induced coupling constants, and its potential
impact on black hole evaporation, has been investigated in
\cite{DS2}. Finally an extension including torsion has been given
in \cite{DS3}.


\begin{center}$\;$
\section{Heat-kernel method}
\end{center}
\noindent According to the idea of quantum vacuum induced
interactions, dynamics of gravity (possibly, including also dark
energy) and gauge interactions emerges from dominant contributions
to the one-loop effective action of non-self-interacting matter
fields coupled to these interactions. In the framework
of the Schwinger proper-time method, the expected terms for
``cosmological constant'' (dark energy), gravity and gauge
interactions can be extracted from the 0th, 1st and 2nd
coefficient of the Seeley--DeWitt heat-kernel expansion,
respectively. In minkowskian signature
\cite{Birrell},\cite{DeWitt}
\begin{align}
 S_{\rm eff}=i\kappa\log\det\mathcal{D}=i\kappa\mathrm{Tr}\log\mathcal{D}=-i\kappa\int\frac{\mathrm{d}s}{s}\;
 \mathrm{Tr}\;e^{-is\mathcal{D}},
 \label{eq:EffectiveAction1}
\end{align}
where $\mathcal{D}$ is an appropriate second-order differential
operator, and $\kappa$ depends on the kind of the ``matter'' field
(its statistics, in principle). E.g.\ for a scalar mode,
$\kappa=\frac{1}{2}$. Making use of the Seeley--DeWitt heat-kernel
expansion in four dimensions,
\begin{align}
\mathrm{Tr}\;e^{-is\mathcal{D}}=\frac{1}{16{\pi}^{2}\left(is\right)^2}\left[A_{0}+A_{1}\left(is\right)+A_{2}\left(is\right)^2+\cdots\right],
\label{eq:Seeley-DeWiitExpansion1}
\end{align}
where $A_{n}$ is the $n$th Seeley--DeWitt coefficient, and next
imposing appropriate cutoffs, i.e.\ an UV cutoff $\varepsilon$ for
$A_0$, $A_1$ and $A_2$, and an IR cutoff $\Lambda$ for $A_2$, we
obtain
\begin{align}
S_{\rm
eff}=\frac{\kappa}{16{\pi}^{2}}\left(\frac{1}{2}A_{0}\varepsilon^{-2}+A_{1}\varepsilon^{-1}+A_{2}\log\frac{\Lambda}{\varepsilon}+\cdots\right).
\label{eq:Seeley-DeWiitExpansion2}
\end{align}
Collecting contributions from various modes, we get the following
lagrangian densities:
\begin{align}
\mathcal{L}_{0}=\frac{1}{64{\pi}^{2}}\varepsilon^{-2}\left(N_{0}-2N_{\frac{1}{2}}+2N_{1}\right),
\label{eq:LagrangeDensity0}
\end{align}

\begin{align}
\mathcal{L}_{1}=-\frac{1}{192{\pi}^{2}}\varepsilon^{-1}\left(N_{0}+N_{\frac{1}{2}}-4N_{1}\right)R,
\label{eq:LagrangeDensity1}
\end{align}
and
\begin{align}
\mathcal{L}_{2}=\frac{1}{384{\pi}^{2}}\log\frac{\Lambda}{\varepsilon}\left(N_{0}+4N_{\frac{1}{2}}\right)\mathrm{tr}F^{2}
\label{eq:LagrangeDensity2}
\end{align}
(see, Table 2 in Appendix for the origin of the numeric coefficients),
 where:
\begin{equation}
\begin{split}
&N_0=\mbox{number of minimal scalar degrees of freedom (dof),}\\
&N_{\frac{1}{2}}=\mbox{number of two-component fermion fields}=\mbox{half the number of fermion dof,}\\
&N_{1}=\mbox{number of gauge fields}=\mbox{half the number of
gauge dof.}
\end{split}
\label{Notations}
\end{equation}
The lagrangian densities $\mathcal{L}_{0}$, $\mathcal{L}_{1}$ and
$\mathcal{L}_{2}$ correspond to the terms $A_{0}$, $A_{1}$ and
$A_{2}$ in \eqref{eq:Seeley-DeWiitExpansion2}, and yield the
cosmological constant, Einstein's gravity and gauge interactions,
respectively. (Here $R$ is the scalar curvature, and $F$ is the
strength of a gauge field, see, the definition \eqref{eq:A.1}.)
Higher-order terms are in principle present (even in classical
case), but they are harmless in typical situations because of
small values of the coefficients following from the cutoffs. An exception appears and is discussed in Sect.\
3.2.

The infamous cosmological constant directly follows from
Eq.~\eqref{eq:LagrangeDensity0} but it is unacceptable in this form because its
value is too huge \cite{Weinberg}, i.e.\ it is at least $10^{120}$ times
greater than expected. Therefore, $A_0$ could, in principle, spoil
the idea of induced interactions but it is not necessarily so. It
appears \cite{Broda} that it is, in principle, possible to tame
the expression \eqref{eq:LagrangeDensity0}, so preserving the
concept of induced interactions consistent.

In the following sections we will consider induced gravity and
standard model gauge interactions.

\begin{center}$\;$\section{Induced gravity}\end{center}
\subsection{Standard approach}

\noindent Assuming the commonly being used, standard, planckian
value of the UV cutoff, $\varepsilon=G$ ($=$\;the Newton
gravitational constant), we directly obtain from
\eqref{eq:LagrangeDensity1}
\begin{align}
\mathcal{L}_{1}=-\frac{1}{12\pi}\left(N_{0}+N_{\frac{1}{2}}-4N_{1}\right)\frac{1}{16\pi
G}R. \label{eq:LagrangeDensity3}
\end{align}

Intuitively, the (quantum) planckian cutoff can be explained as
following from a classical gravitational cutoff imposed by the
black hole horizon. Namely, the description of matter in terms of
particles, or even the notion of particles itself, is not valid
for particles of enormous, i.e.\ planckian, masses because of the
mechanism of black-hole formation (see, Fig.~1). We shall return
to this thread in the next subsection.

The result \eqref{eq:LagrangeDensity3} has been already explicitly
presented in \cite{Akama} (there is a misprint in the coefficient
in front of $N_1$ of his final formula (4.2)). It has been also
rederived in the framework of the heat-kernel method, in the
context of supersymmetry, in \cite{Tanaka}. Finally,
Eq.~\eqref{eq:LagrangeDensity3} can also be easily recovered from
the data given in \cite{Visser}. The aim of the former two papers
was to investigate the influence of gauge fields on the sign of
$\mathcal{L}_{1}$. Obviously, the gauge fields tend to change the
sign of $\mathcal{L}_{1}$. Instead, our aim, in this subsection,
is to emphasize that Eq.~\eqref{eq:LagrangeDensity3} directly
yields a realistic value of the Newton gravitational constant,
provided the planckian UV cutoff is given. Namely, putting, e.g.,
$N_{0}=0$, $N_{\frac{1}{2}}=45$ and $N_{1}=0$ in
\eqref{eq:LagrangeDensity3} we get in front of the standard
Hilbert--Einstein action $\frac{45}{12\pi}\approx
1.19=\mathcal{O}(1)$, which is an impressive coincidence. Taking
into account an approximate character of the derivation such a
high precision is absolutely unnecessary and seems to be rather
accidental. Therefore, assuming non-zero $N_{0}$ and $N_{1}$,
e.g.\ $N_{0}=4$ and $N_{1}=12$, yields $\frac{1}{12\pi}\approx
0.03$, and it is perhaps less impressive but phenomenologically
acceptable as well. The proposed value of $N_{\frac{1}{2}}$
corresponds to $3\times(3+3\times4)=45$ fermion two-component
field species contained in the standard model (3 families of
leptons and quarks in 3 colors). In the second example, we admit
the existence of the Higgs scalar, $N_0=4$, and the contribution
of $N_1=1+3+8=12$ gauge fields. Since the gauge fields themselves
are also induced entities, their contribution to the count is
disputable. The existence of the Higgs particle itself is
disputable as well.

\subsection{Alternative (cutoff independent or self-cutoff) approach}
\noindent Strictly speaking, the whole approach presented in the
previous subsection, and being in accordance with a standard,
commonly accepted point of view, is not quite logically
consistent. The lack of the full logical consistency is a
consequence of the fact that the assumed planckian value of the UV
cutoff, in principle, follows from the value of the (effective) Newton
gravitational constant that is just being induced. In other words,
a consistent reasoning should be independent of any explicit value
of the UV cutoff. Of course, it is impossible to derive the
(numeric) value of $G$ or, equivalently, of the UV cutoff in such
a framework, but nevertheless some physically non-trivial
conclusions can be drawn.

First of all, we should somehow explain the appearance of the
Planck scale in physics. Apparently, there are the two main points
of view in this respect. The first point of view, due to Planck
himself, appeals to a possibility to construct an appropriate
dimensionfull quantity out of several fundamental physical
constants. It is theory independent and natural for simple
dimensional grounds, but it lacks a firm physical support.
Besides, that approach is not able to yield any purely numeric
coefficient. The second approach instead tries to derive the
Planck scale using some physical, theory grounded arguments. That
approach dates back to the papers
\cite{Peres},\cite{Mead},\cite{Mead1}, and conforms to our point
of view. Roughly, the idea, one could call self-cutoff induced
gravity, consists in identification of the Schwarzschild diameter
and the Compton wavelength. Naively, the reasoning according to
these guidelines could look like follows. Literally repeating the
derivation of \eqref{eq:LagrangeDensity3}, but this time with an
unpredefined UV cutoff $M$ (in mass units), we get
\begin{align}
\mathcal{L}_{1}=-\frac{M^{2}}{16\pi}\cdot\frac{N}{12\pi}R,
\label{eq:LagrangeDensity4}
\end{align}
where $N$ is the ``effective number'' of particle species, e.g.\
$N=N_{0}+N_{\frac{1}{2}}-4N_{1}$ (see
\eqref{eq:LagrangeDensity3}). Obviously, the Schwarzschild
solution of the Einstein equation following from
\eqref{eq:LagrangeDensity4} is independent of the coefficients in
\eqref{eq:LagrangeDensity4}, and it assumes the well-known form
\begin{align}
{ds}^2=\left(1-\frac{\mu}{r}\right){dt}^2-\left(1-\frac{\mu}{r}\right)^{-1}{dr}^2-r^{2}{d\Omega}^2,
\label{eq:SchwarzschildSolution1}
\end{align}
where $\mu$ is an as yet undefined parameter. The (standard)
linearized and well-known version of the Einstein equation for the
metric \eqref{eq:SchwarzschildSolution1} (Newtonian limit) with a
point-like source representing a particle of the highest
admissible mass $M$ (UV cutoff) with coefficients of
Eq.~\eqref{eq:LagrangeDensity4} reads
\begin{align}
\frac{M^{2}}{16\pi}\cdot\frac{N}{12\pi}\cdot\frac{1}{2}\;\Delta\left(-\frac{2\mu}{r}\right)=\frac{1}{2}M\delta^{3}(\mathbf{r}),
\label{eq:SchwarzschildSolution2}
\end{align}
where $\Delta$ is the three-dimensional laplacian, and
$\delta^3(\mathbf{r})$ is the Dirac delta. From
\eqref{eq:SchwarzschildSolution2}, it follows that
\begin{align}
MN\mu=24\pi. \label{eq:SchwarzschildSolution3}
\end{align}
On the other hand, the value of the Compton wavelength for the
particle of the mass $M$ is
$$\lambda_c=\frac{2\pi}{M}.$$
Equating $\lambda_c$ and the Schwarzschild diameter
\begin{align}
2r_s=2\mu, \label{eq:Miu Equation}
\end{align}
we obtain from \eqref{eq:SchwarzschildSolution3} the final
(dimensionless) result:
\begin{align}
N=24. \label{eq:SchwarzschildSolution4}
\end{align}

Typically $r_{s}<<\lambda_{c}$ and therefore no gravity concepts
enter quantum theory discussion. But when the mass $M$ grows,
$r_{s}$ grows linearly, whereas $\lambda_{c}$ decreases. When the
both values become comparable the particle can be intuitively
considered as trapped in its own black hole (Fig.\ 1).
\begin{figure}\label{Fig.1}
\begin{center}
\includegraphics[scale=1.2]{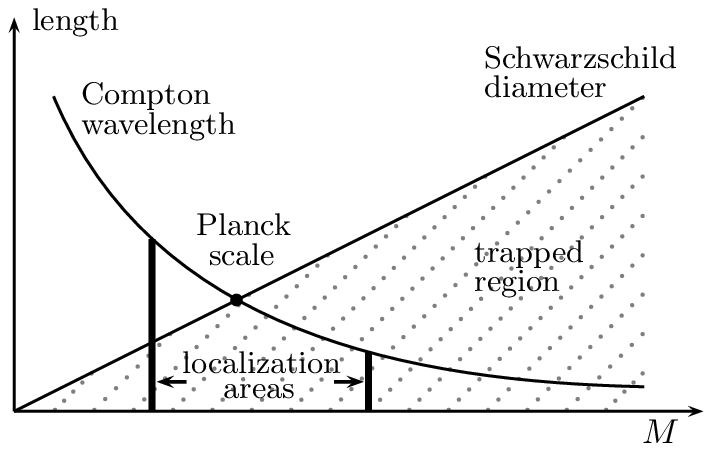}
\end{center}
\begin{quote}
\textit{\textbf{Fig.\ 1}: Qualitative picture of the emergence of
the Planck scale. The left localization area corresponds to the
regime of standard particle-field theory formalism, whereas the
right one is outside the scope of this formalism.}
\end{quote}
\end{figure}

We would like to emphasize that the constraint
\eqref{eq:SchwarzschildSolution4} which is qualitatively fully
consistent with our earlier ones (derived in Sect.\ 3.1) is
derived owing to the self-cutoff assumption but not in general
induced gravity. In this place, we could hastily conclude that the
focus is now shifted to analyzing the role of the ``effective
number'' of degrees of freedom of matter fields (see
Eq.~\eqref{eq:SchwarzschildSolution4}). Unfortunately, such a
conclusion would be not quite correct because there is a problem
with higher-order Seeley--DeWitt coefficients. They are harmless
for small Riemann curvature $R$, but when the Riemann curvature
$R$ is of the order of $M^2$, an infinite tower of
$\left(R/M^2\right)^n$ corrections to the Einstein term appears
($R$ denotes here not only the scalar curvature but symbolically
all kinds of curvature terms of the corresponding dimension). This
could, in principle, invalidate the whole argumentation yielding
the result \eqref{eq:SchwarzschildSolution4}. Therefore, we claim
that the proper conclusion to be drawn in the end of this Section
is as follows. The derivation is consistent, and the
phenomenologically reasonable result $N\approx24$ is obtained
provided the formula for the Schwarzschild radius \eqref{eq:Miu
Equation} is stable against the influences of the infinite tower
of higher-curvature corrections.

Strictly speaking, we should expect some form of a generalization
of the Schwarzschild solution and of the Schwarzschild radius. The
well-known linear relation between the radius of the event horizon
and the mass \eqref{eq:Miu Equation} follows from the form of the
Schwarzschild solution of the Einstein equation according to the
argumentation presented between Eq.~\eqref{eq:LagrangeDensity4}
and Eq.~\eqref{eq:SchwarzschildSolution2}. It is true for
$\left|R_{H}\right|<<M^2$ (the curvature $R$ at the event horizon
$H$ is much less than the UV cutoff), otherwise the
Hilbert--Einstein action is modified by higher-order terms in $R$,
and the classical result could be invalidated. But it appears that
phenomenologically the ``effective number'' of particle species
$N$ qualitatively conforms to \eqref{eq:Miu Equation}. This result
confirms that the derivation is somehow insensitive to terms of
higher-order in $R$. Such a kind of the result could be of
interest in application to mini black holes where, in principle,
one should not discard higher-order terms.

\begin{flushleft}$\:$\subsection{Entropy}\end{flushleft}
\noindent In this subsection, we would like to draw reader's
attention to an independent argument in favour of the idea of
induced gravity. Namely, we will show that not only the action but
also the entropy sums up appropriately, i.e.\ gravitational
entropy of a black hole can be recovered from entropies of
``fundamental'' fields. In other words, what applies to the
actions remains in force in the case of the entropies. A slight
complication follows from the fact that the notion of the entropy
is not quite unique. In principle, in the context of gravity usually two
notions of the entropy appear: so-called, ``geometrical entropy'' and
``thermodynamical entropy''. Since the derivation of the geometrical
entropy \cite{Larsen Wilczek} uses the heat-kernel method, the
result for the entropy is analogous to the previous one for the
action. Namely, in close analogy to
Eq.~\eqref{eq:LagrangeDensity3} we have
\begin{align}
S_{g}=\frac{1}{12\pi}\left(N_{0}+N_{\frac{1}{2}}-4N_{1}\right)S_{A},
\label{eq:Entropy1}
\end{align}
where $S_{g}$ denotes the geometrical entropy and $N_{0}$,
$N_{\frac{1}{2}}$, $N_{1}$ are defined in \eqref{Notations}. Here,
the black hole entropy
\begin{align}
S_{A}=\frac{1}{4G}A, \label{eq:Entropy2}
\end{align}
where $A$ is the area of the black hole horizon. Actually, we
reproduce the same bound as that given in
\eqref{eq:SchwarzschildSolution4},
provided the expected value of $S_g=S_A$.

It appears that the approach making use of the thermodynamical
entropy yields a slightly other result. Now, we have \cite{Li
Zhong-hang}
\begin{align}
S_{B}=\frac{1}{90\pi}S_{A}, \label{eq:Entropy3}
\end{align}
and
\begin{align}
S_{F}=\frac{7}{16}\cdot\frac{1}{90\pi}S_{A}, \label{eq:Entropy4}
\end{align}
for a bosonic and a fermionic degree of freedom, respectively.
Therefore, the final formula reads
\begin{align}
S_{t}=\frac{1}{90\pi}\left(N_{B}+\frac{7}{16}N_{F}\right)S_{A},
\label{eq:Entropy5}
\end{align}
where $N_{B}$ and $N_{F}$ is the number of bosonic and fermionic
degrees of freedom, respectively. In terms of $N_{0}$,
$N_{\frac{1}{2}}$, $N_{1}$ we could rewrite \eqref{eq:Entropy5} as
(see, Table 2 and further formulas in Appendix)
\begin{align}
S_{t}=\frac{1}{90\pi}\left(N_{0}+\frac{7}{8}N_{\frac{1}{2}}+2N_{1}\right)S_{A}.
\label{eq:Entropy6}
\end{align}
Evidently, the formula \eqref{eq:Entropy6} differs from
\eqref{eq:Entropy1} but nevertheless qualitatively the result is
essentially the same as earlier for the considered combinations of field species.

In the end of this subsection, we would like to stress that the
presented observations are not new, only the point of view is
changed. From traditional point of view, the most natural way to
explain the origin of the black hole entropy is to treat it as
``entanglement entropy'' for constituent fields \cite{Jacobson}.
Accordingly, Eq.~\eqref{eq:Entropy1} is interpreted as a
derivation of $S_g$. From our point of view the, so-called,
``species problem'', signaled in \cite{Jacobson},  does not exist.

\begin{center}$\;$
\section{Induced gauge interactions}
\end{center}
\noindent In this section, we will concentrate on the possibility
of quantum generation of gauge interactions in the context of the
standard model. From technical point of view, we will be
interested in the second Seeley--DeWitt coefficients for
appropriate matter fields. The corresponding term has been already
given in \eqref{eq:LagrangeDensity2} but now we would like to
adapt it to the context of the standard model. Adopting the matter
contents of the lagrangian of the standard model we display all
contributions to the respective gauge parts in Table 1.
\begin{figure}\label{Table 1}
\includegraphics[scale=0.95]{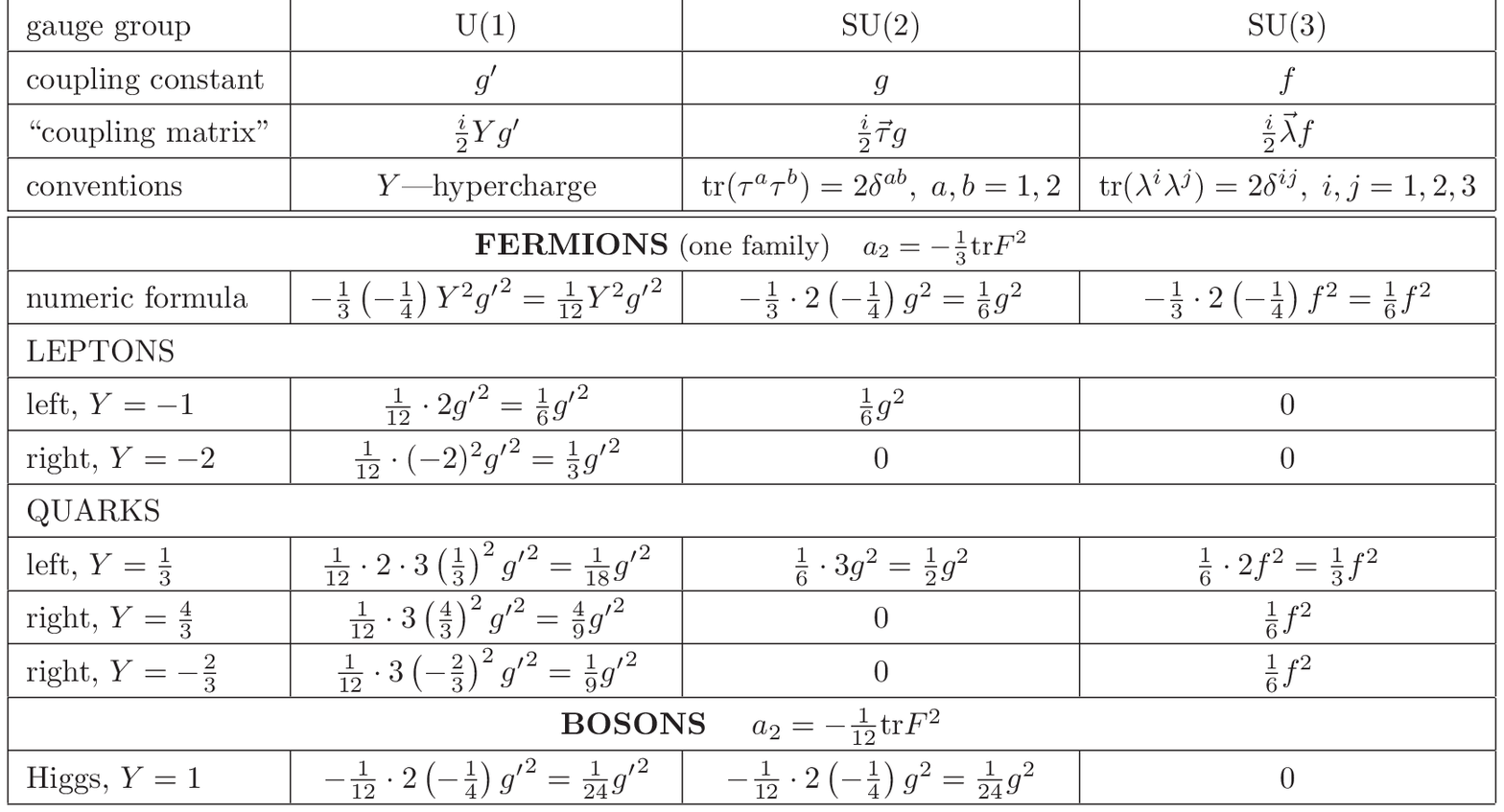}
\begin{quote}
\textit{\textbf{Table 1}: Contributions to induced gauge field
coupling constants coming from various ``matter field'' species in the
framework of the standard model.}
\end{quote}
\end{figure}

Here the assumed implicit convention for the operator of covariant
derivative is
\begin{align}
D_{\mu}=\partial_{\mu}+\vec{X}\cdot\vec{A}_{\mu},
\label{eq:Covariant1}
\end{align}
where $\vec{X}$ is the ``coupling matrix'' given in the third row
of Table 1. More precisely, $\vec{X}$ is a tensor product with two
matrix units corresponding to the other two gauge groups, yielding
additional coefficients, 2 or 3. In principle, the coefficients
given in each column and multiplied by
\begin{align}
\frac{1}{16{\pi}^2}\log\frac{M}{m}, \label{Coefficient1}
\end{align}
where $M$ and $m$ is an UV and an IR cutoff, respectively, in mass
units, should sum up to $\frac{1}{4}$, a standard normalization
term in front of $F^2$. More generally, we have the following
theoretical bound:
\begin{align}
\frac{{g_{i}}^2}{16{\pi}^2}\sum_{n}\alpha_{(i)n}\log\frac{M_n}{m_n}=\frac{1}{4},
\label{Coefficient2}
\end{align}
where $g_i\;(i=1,2,3)$ is one of the three coupling constants,
$\alpha_{(i)n}$ are corresponding numeric coefficients from Table
1, and the sum concerns all matter fields. We can confidently set
$M_n=M_{\rm P}$ (Planck mass), but the choice of $m_n$ is less obvious.
Fortunately, the logarithm is not very sensitive to a change of
the argument.

Now, the data given in Table 1 can be used to reproduce a number
of phenomenologically realistic results. Assuming for simplicity
(or as an approximation) fixed values of $M_n$ and $m_n$ for all
species of matter particles, we can uniquely rederive following
\cite{Terazawa} the Weinberg angle $\theta_{\rm \textsc{w}}$,
\begin{align}
\sin^2\theta_{\rm
\textsc{w}}=\frac{{g'}^2}{g^2+{g'}^2}\approx0.38\;.
\label{eq:WeinbergAngle}
\end{align}
Unfortunately, estimation of the coupling constants requires definite values
of infrared cutoffs $m_n$. Anyway, for $m_n$ of the order of the
mass of lighter particles of the standard model we obtain
\begin{align}
\alpha=\frac{e^2}{4\pi}=\frac{g^2\sin^2\theta_{\rm
\textsc{w}}}{4\pi}=\mathcal{O}(0.01), \label{eq:SubtleConstant}
\end{align}
and
\begin{align}
g=f=\mathcal{O}(1),
 \label{eq:Coupling}
\end{align}
which is phenomenologically a very realistic estimate.

Alternatively, the bound \eqref{eq:SchwarzschildSolution4} can
give some, for example, (non-unique) limitations on the ratio of the two scales
$M$ and $m$, provided the scale of interactions
$g=f=\mathcal{O}(1)$ is assumed.

\begin{center}$\;$
\section{Final remarks}
\end{center}
\noindent In this paper, we have presented a number of arguments
supporting the idea of the old-fashioned ``one-loop dominance''
version of induced gravity and gauge interactions in the spirit of
Sakharov. All coupling constants of fundamental gauge
interactions, including gravity, have been shown to assume
phenomenologically realistic values, provided the planckian value
of the UV cutoff is given. Besides the action, also the black hole
entropy has been shown to fit this picture. As another, UV cutoff
free interpretation of the consistency of induced gravity, an
estimate of the generalized Schwarzschild radius has been proposed
in the framework of an idea of self-cutoff induced gravity.

It seems that the brane induced gravity could be probed with the
approach presented. For example, in five dimensions, at least
formally, we would have got ${M_{*}}^{3}$ instead of $M^{2}$ in
front of the five-dimensional version of
Eq.~\eqref{eq:LagrangeDensity4}. In this case some difficulty
could follow from the fact that the matter fields gravity is
supposed to be induced from do not live in higher dimensions. But
this subject is outside the scope of our paper.
\eject
\section*{\begin{center} Acknowledgments\end{center}}
We are grateful to many persons for their critical and fruitful
remarks which we have utilized in this paper. This work was
supported in part by the Polish Ministry of Science and Higher
Education Grant PBZ/MIN/008/P03/2003 and by the University of
{\L}\'od\'z grant.

\section*{\begin{center} Appendix\\Seeley--DeWitt and entropy coefficients \end{center}}
For reader's convenience we present below (in Table 2) the
Seeley--DeWitt (``Hamidew'') coefficients and the entropy
coefficients used (except $k_2$ for a massless vector) in the main
text.
In the terminology of Misner, Thorn and Wheeler our sign
convention corresponds to the Landau--Lifshitz timelike one, i.e.\
the metric signature is $\left(+---\right)$ and
$R^{\alpha}_{\;\beta\gamma\delta}=\partial_{\gamma}\Gamma^{\alpha}_{\;\beta\delta}-\cdots$.
Our conventions concerning gauge fields are as follows:
\begin{equation} \tag{A.1}
\begin{split}
&D_{\mu}=\nabla_{\mu}+A_{\mu},\\
&F_{\mu\nu}=\partial_{\mu}A_{\nu}-\partial_{\nu}A_{\mu}+\left[A_{\mu},A_{\nu}\right].
\end{split}
\label{eq:A.1}
\end{equation}
\\

\begin{center}
 \begin{tabular}{|c||c|c|c||c|}
  \hline
  particle & \multicolumn{3}{c||}{Seeley--DeWitt coefficients} &
  entropy\\
  \cline{2-4}
   & \;$k_0$\;  & \;\;\;\;$k_1$\;\;\;\; & $k_2$ & coefficient $l$\\
  \hline \hline
  minimal scalar & 1 & $\frac{1}{6}$\; \cite{DeWitt} &
  $\frac{1}{12}$\;
  \cite{DeWitt}& 1\; \cite{Li Zhong-hang}\\
  \hline
  Weyl spinor & 2 & $-\frac{1}{6}$\; \cite{DeWitt} & $-\frac{1}{3}$\;
  \cite{DeWitt}& $\frac{7}{8}$\; \cite{Li Zhong-hang}\\
  \hline
  massless vector & 2 & $-\frac{2}{3}$\; \cite{Birrell} & $-\frac{11}{24}$\;
  \cite{Vassilevich} & 2\; \cite{Li Zhong-hang}\\
  \hline
 \end{tabular}
\end{center}
\begin{quote}
\textit{\textbf{Table 2}: Seeley--DeWitt coefficients and entropy
coefficients. In brackets, we have given the references where the
coefficients can be found explicitly or almost explicitly (i.e.\
after few-minute calculations).}
\end{quote}
We have assumed the following notation:
\begin{equation} \tag{A.2}
\begin{split}
&a_{0}(x)=k_{0},\\
&a_{1}(x)=-k_{1}R,\\
&a_{2}(x)=k_{2}\,{\rm{tr}}F^{2}+k'_{2} \textrm{``curvature
terms''},
\end{split}
\label{eq:A.2}
\end{equation}
and
\begin{equation} \tag{A.3}
S_{t}=l\cdot \frac{1}{90\pi}\cdot\frac{A}{4G}. \label{eq:A.3}
\end{equation}
Interested reader can find $k'_2$ in \cite{Birrell}, \cite{Visser}
or \cite{Vassilevich}.


\newpage


\begin{thebibliography}{}
\bibitem[1]{Sakharov}A.\ D.\ Sakharov, ``Vacuum Quantum Fluctuations
in Curved Space and the Theory of Gravitation'', Gen.\ Rel.\
Grav.\ \textbf{32}, 365--367 (2000), translated from Dokl.\ Akad.\
Nauk SSSR \textbf{170}, 70–71 (1967).

\bibitem[2]{Zeldovich}Y.\ B.\ Zel’dovich, ``Interpretation of Electrodynamics as a Consequence of
Quantum Theory'', Pis’ma ZhETF \textbf{6}, 922--925 (1967); [JETP
Lett.\ \textbf{6}, 345--347 (1967)].

\bibitem[3]{Bjorken}
J.\ D.\ Bjorken, ``A Dynamical Origin for the Electromagnetic
Field'', Annals Phys.\ \textbf{24}, 174--187 (1963).

\bibitem[4]{Birula}
I.\ Bia{\l}ynicki-Birula, ``Quantum Electrodynamics without
Electromagnetic Field'', Phys.\ Rev.\ \textbf{130}, 465--468
(1963).

\bibitem[5]{Terazawa1}
H.\ Terazawa, Y.\ Chikashige and K.\ Akama,  ``Unified Model of
the Nambu--Jona-Lasinio Type for all Elementary-Particle
Forces'', Phys.\ Rev.\ D \textbf{15}, 480--487 (1977).

\bibitem[6]{Akama1}
K.\ Akama, Y.\ Chikashige and T.\ Matsuki, ``Unified Model of the
Nambu--Jona-Lasinio Type for the Gravitational and
Electromagnetic Forces'', Prog.\ Theor.\ \textbf{59}, 653--655
(1978).


K.\ Akama, Y.\ Chikashige, T.\ Matsuki and H.\ Terazawa, ``Gravity
and Electromagnetism as Collective Phenomena of
Fermion-Antifermion Pairs'',  Prog.\ Theor.\ \textbf{60}, 868--877
(1978).

\bibitem[7]{Akama2}
K.\ Akama, ``An Attempt at Pregeometry---Gravity with Composite
Metric'', Prog.\ Theor.\ Phys.\ \textbf{60}, 1900--1909 (1978).

\bibitem[8]{Terazawa2}
H.\ Terazawa, Y.\ Chikashige, K.\ Akama and T.\ Matsuki,  ``Simple
Relation Between the Fine-Structure and Gravitational Constants'',
Phys.\ Rev.\ D \textbf{15}, 1181--1183 (1977).

\bibitem[9]{Landau}
L.\ D.\ Landau, ``On the Quantum Theory of Fields, in Niels Bohr
and the Development of Physics'' (Essays dedicated to Niels Bohr
on the occasion of his seventieth birthday, Edited by W.\ Pauli),
(New York: McGraw-Hill Book Co., 1955), p.\ 52--69.

\bibitem[10]{Akama3}
K.\ Akama, ``Pregeometry'', in Lecture Notes in Physics, 176,
Gauge Theory and Gravitation, Proceedings, Nara, (1982), edited by
K.\ Kikkawa, N.\ Nakanishi and H.\ Nariai, (Springer--Verlag,
1983), p.\ 267--271; [arXiv:hep-th/0001113].

\bibitem[11]{Akama4}
K.\ Akama, ``Compositeness Condition at the Next-to-Leading Order
in the Nambu--Jona-Lasinio Model'', Phys.\ Rev.\ Lett.\
\textbf{76}, 184 (1996).

K.\ Akama and T.\ Hattori, ``Exact Scale Invariance of
Composite-Field Coupling Constants'', Phys.\ Rev.\ Lett.\
\textbf{93}, 211602 (2004).

\bibitem[12]{Visser}M.\ Visser, ``Sakharov's Induced Gravity: a
Modern Perspective'', Mod.\ Phys.\ Lett.\ A \textbf{17}, 977--992
(2002); [arXiv:gr-qc/0204062].

%
\bibitem[13]{DS1}
G.\ Denardo and E.\ Spallucci, ``Induced Quantum Gravity from Heat
Kernel Expansion'', Nuovo Cim.\ A \textbf{69}, 151--159 (1982).

\bibitem[14]{DS2}
G.\ Denardo and E.\ Spallucci,
``Finite Temperature Spinor Pregeometry'',
Phys.\ Lett.\ B \textbf{130}, 43--46 (1983);
``Finite Temperature Scalar Pregeometry''
Nuovo Cim.\ A \textbf{74}, 450--460 (1983).

\bibitem[15]{DS3}
G.\ Denardo and E.\ Spallucci, ``Curvature and Torsion from
Matter'', Class.\ Quantum Grav.\ \textbf{4}, 89--99 (1987).

\bibitem[16]{Birrell}N.\ D.\ Birrell and P.\ C.\ W.\ Davies  ``Quantum Fields in Curved Space'',
(Cambridge University Press, 1982), Chapt.\ 6.

\bibitem[17]{DeWitt}B.\ DeWitt, ``The Global Approach to Quantum Field Theory'',
(Oxford Science Publications, 2003), Chapt.\ 27.

\bibitem[18]{Weinberg}S.\ Weinberg, ``The Cosmological Constant Problem'', Rev.\ Mod.\
Phys.\ \textbf{61}, 1--23 (1989).

\bibitem[19]{Broda}B.\ Broda, P.\ Bronowski, M.\ Ostrowski and M.\ Szanecki, ``Vacuum Driven Accelerated
Expansion'', Ann.\ Phys.\ (Berlin), 1–9 (2008) / DOI
10.1002/andp.200810314; [arXiv:0708.0530].

\bibitem[20]{Akama}K.\ Akama, ``Pregeometry Including
Fundamental Gauge Bosons'', Phys.\ Rev.\ D \textbf{24}, 3073--3081
(1981).

\bibitem[21]{Tanaka}M.\ Tanaka, ``Three Generations or More
for an Attractive Gravity?'', Phys.\ Rev.\ D \textbf{53},
6941--6945 (1996); [arXiv:hep-ph/9504259].

\bibitem[23]{Peres}
A.\ Peres and N.\ Rosen, ``Quantum Limitations on the Measurement
of Gravitational Fields'', Phys.\ Rev.\ \textbf{118}, 335--336
(1960).

\bibitem[24]{Mead}
C.\ A.\ Mead, ``Possible Connection Between Gravitation and
Fundamental Length'', Phys.\ Rev.\ \textbf{135}, B849--B862
(1964).

\bibitem[25]{Mead1}
C.\ A.\ Mead, ``Observable Consequences of Fundamental-Length
Hypotheses'', Phys.\ Rev.\ \textbf{143}, 990--1005 (1966).

\bibitem[26]{Larsen Wilczek}F.\ Larsen and F.\ Wilczek, ``Renormalization of Black Hole Entropy and of the Gravitational Coupling
Constant'', Nucl.\ Phys.\ B \textbf{458}, 249--266 (1996);
[arXiv:hep-th/9506066].

\bibitem[27]{Li Zhong-hang}L.\ Zhong-heng, ``Entropy
of Spin Fields in Schwarzschild Spacetime'', Int.\ J.\ Theor.\
Phys.\ \textbf{39}, 253--257 (2000).

\bibitem[28]{Jacobson}T.\ Jacobson, ``Black Hole Entropy and Induced
Gravity''; [arXiv:gr-qc/9404039].

\bibitem[29]{Terazawa}H.\ Terazawa, K.\ Akama and
Y.\ Chikashige, ``What Are the Gauge Bosons Made of?'', Prog.\
Theor.\ Phys.\ \textbf{56}, 1935--1938 (1976).

\bibitem[30]{Vassilevich}D.\ V.\ Vassilevich, ``Heat Kernel Expansion: User's Manual'', Phys.\ Rept.\ \textbf{388}, 279--360
(2003); [arXiv:hep-th/0306138].

\end{thebibliography}
\end{document}